# LONGITUDE DEPENDANCE OF SOLSTICIAL HADLEY CELL DETECTED AT THE EDGE OF THE MASSIVE MARTIAN ERG

M. Kuassivi
B.P. 170374 Cotonou, Benin

## ABSTRACT

Using public HIRISE images of Mars, I derive the wind directions at high northern lattitudes where many interesting eolian features are observed. Barchan dunes show a prominent wind direction from the north indicating that they formed during the southern summer. But a few record consistent SE winds near the Utopia Planitia basin. This wind reversal is consistent with a local perturbation of the solsticial Hadley cell caused by the geological depression.

## 1. INTRODUCTION

The North—South slope of the martian topography produces a Hadley circulation that is asymmetric about the equator. However, to test this hypothesis, sophisticated models, accounting for all the processes thought to affect the Martian meteorology (topography, surface thermal inertia, surface albedo, solar heating) are needed.

Richardson & Wilson (2002) have shown that the zonal mean component of topography is the dominant factor in causing a lattitude—asymmetric Hadley circulation on Mars. It is now well accepted that the surface height is clearly affecting the atmospheric circulation but further numerical investigations of the influence of topography require huge computing power (higher resolution). If the role of elevated regions has been intensively investigated, it is not yet the case for the lowlands.

The northern lowlands is the site for hypothesized standing bodies of water in past Mars history. It is occupied by a widespread unit: the Vastitas Borealis Formation (V.B.F.). Within the V.B.F. the Utopia Planitia centered at 50°N 118° E is the largest impact basin and lies 4—5 km below the mean radius of the planet. Interestingly enough, it is also the Martian region where the Viking 2 lander touched down and began exploring on September 3, 1976.

In this paper, I argue that eolian features observed at the edge of Utopia Planitia basin show imprints of Hadley cell perturbation.

## 2. METHODOLOGY AND OBSERVATIONS

Near the north polar cap of Mars, the landscape is dominated by sand dunes forming a massive erg. At the edge of this erg, the sand is in shorter supply and many barchan type dunes form in sand starved areas. Barchans are crescent-shaped dunes with a steep slip face bordered by two horns oriented downwind. They form by uni—directional winds and so when seen, give a clear indication of wind direction (Bagnold, 1941).

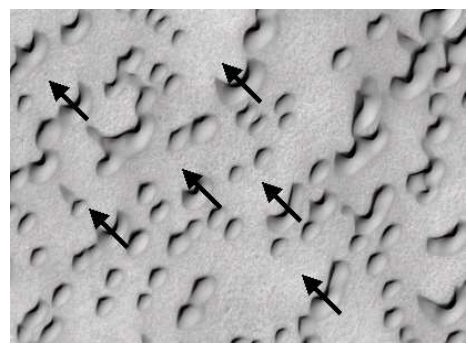

**Fig 1.** A HIRISE image of barchans dunes (PSP_001608_2560) with their interpreted directions surimposed on the image.

| Lat | Long (east) | Obs | Wind |
|---|---|---|---|
| +78.7° | 71.2° | ESP_016113_2590 | -80° |
| +76.1° | 98.0° | ESP_018994_2840 | +140° |
| +75.8° | 94.1° | ESP_019211_2560 | +140° |
| +76.2° | 95.4° | ESP_018525_2565 | +140° |
| +76.2° | 95.4° | ESP_018011_2565 | +140° |
| +75.9° | 94.0° | ESP_017926_2840 | +140° |
| +76.0° | 86.9° | PSP_001608_2560 | +140° |
| +78.5° | 118.9° | ESP_019112_2815 | -145° |
| +76.6° | 104.1° | ESP_018815_2570 | +150° |
| +76.6° | 104.1° | ESP_018229_2830 | +150° |
| +76.7° | 109.6° | PSP_001660_2570 | +150° |
| +80.0° | 122.5° | ESP_019368_2600 | +150° |
| +79.4° | 169.3° | PSP_007064_2595 | -70° |
| +77.4° | 210.1° | ESP_017387_2575 | -90° |
| +76.9° | 226.6° | ESP_017426_2570 | -130° |
| +78.1° | 240.0° | ESP_017531_2580 | -120° |
| +76.4° | 231.6° | PSP_009917_2835 | -100° |
| +78.0° | 233.8° | PSP_010015_2580 | -120° |
| +79.3° | 245.5° | ESP_019469_2595 | +90° |
| +77.7° | 256.7° | ESP_017715_2580 | -110° |
| +78.1° | 251.0° | PSP_008274_2580 | +100° |
| +76.1° | 270.6° | PSP_009394_2565 | -100° |
| +75.3° | 293.3° | ESP_019000_2845 | -90° |
| +76.5° | 297.4° | ESP_018742_2565 | -90° |
| +73.8° | 286.4° | PSP_010422_2565 | -100° |
| +76.5° | 297.4° | PSP_001930_2565 | -90° |
| +77.5° | 300.1° | ESP_017898_2575 | +170° |
| +73.7° | 317.0° | ESP_016447_2540 | -70° |
| +77.5° | 300.1° | PSP_009736_2575 | +170° |
| +77.5° | 300.1° | PSP_009696_2575 | +160° |
| +77.5° | 300.1° | PSP_008839_2575 | +170° |
| +75.9° | 314.7° | PSP_001758_2560 | -90° |

**Table.1** Datasets and wind directions.

The public HIRISE image data are revieweved for the presence of barchan dunes in between latitude 60° and 80°. Such landforms appear in at least 32 images (see table 1.).

Images are interpreted by eye and the angle of each eolian feature is recorded and averaged to produce the dominant wind direction of the imaged region (see Fig. 1).

3.  CONCLUSION

The angular coverage of this analysis is sufficient to observe a prominent wind direction from the north indicating that the features formed during the southern hemisphere summer, when the strongest wind blows that direction (Thomas, 1981). However, South—East winds are observed over a large region in between 98°E and 123°E. This reversal is well correlated with the northern frontier of the Utopia Planitia basin. Fig. 2 shows the elevation map obtained with the Mars Orbiter Laser Altimeter (M.OL.A.) on which the wind directions have been surimposed.

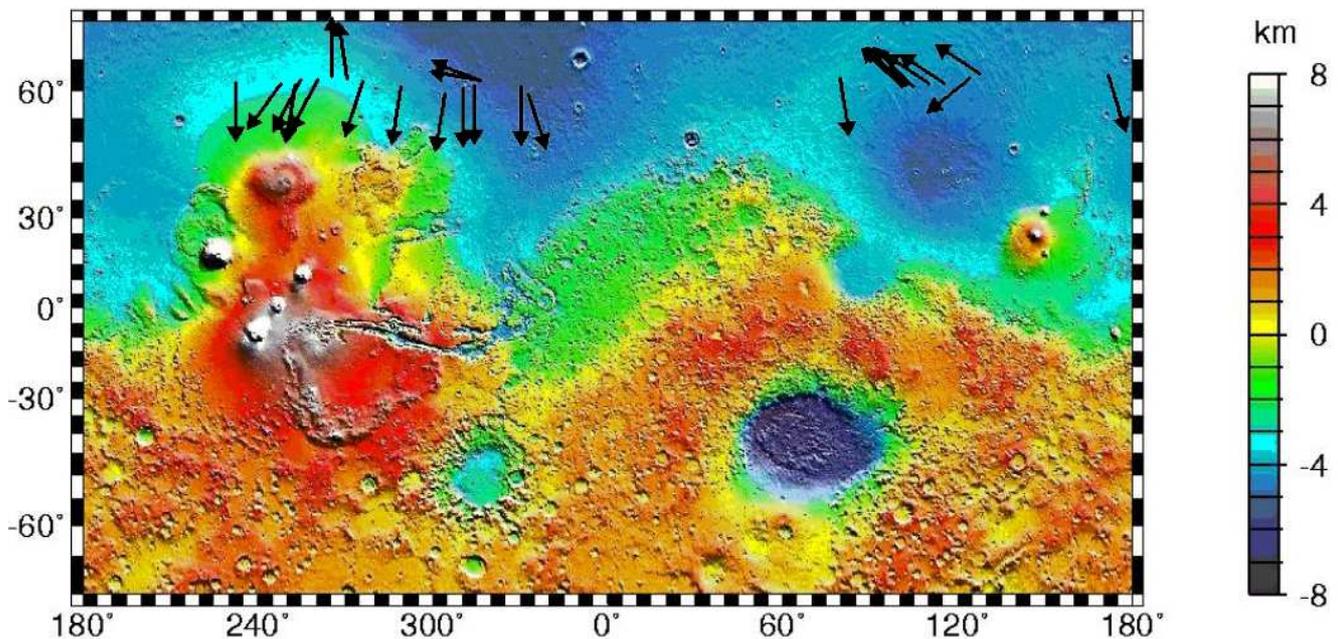

**Fig. 2** M.O.L.A. elevation map of Mars with the wind directions surimposed.

Thermal forcing that depends on the surface height is expected to occur both on Earth and on Mars. Because atmosphere is warmer over an elevated surface, the peak heating shifts towards higher topography. The latitude of the dividing streamline shifts in the same direction.

When looking at the elevation map of Mars (Fig. 2), it is clear that the shift is southward at most longitudes (Richardson & Wilson, 2002). However, following the same argument, one expects the shift to be northward near the Utopia Planitia basin. Such a shift may result in wind reversal.

Comprehensive general circulation of the Martian atmosphere is needed to investigate the seasonal changes over the surface that may be related to the occurence of life. Moreover, the design of missions on Mars requires a good knowledge of the Martian environment and its variability (for aerobraking, aero-capture, descent and landing, instrument specification). As a matter of facts, the analysis of the longitudinal dependance of the martian hadley cell may reveal a handy tool for such investigations.

This work is based on NASA/JPL/ Arizona University images made available for the public by the HIRISE team.